\documentclass[12pt,letterpaper]{article}
\usepackage[utf8]{inputenc}
\usepackage[T1]{fontenc}
\usepackage{amssymb}
\usepackage{graphicx}
\usepackage[left=1.50in, right=1.50in, top=1.50in, bottom=1.00in]{geometry}
\author{John M.~LoSecco\\University of Notre Dame du Lac}
\title{Gravitational Lensing of Supernova Neutrino Bursts}
\date{\today}
\begin{document}
\maketitle

\begin{abstract}
  Supernova neutrino bursts have been observed from extragalactic distances.
  This note addresses the question
  of how gravitational lensing could distort
  the information in the burst.  We apply the gravitational
  lens hypothesis to try to understand the time and brightness structure of the SN1987A neutrino observations.
  Estimates of a possible lensing mass and alignment are made.
  These estimates suggest a path to verification.
\end{abstract}


\maketitle
%
\section{Introduction}
Observations of neutrinos from gravitational stellar collapse, a supernova,
has the ability to peer into the heart of a transient critical event in the
evolution of our universe.  In a few seconds a mature massive star transforms
into a neutron star or black hole releasing most of the binding energy as
neutrinos.  The subsequent explosion ejects the heavy elements into
interstellar space and creates those beyond the nuclear binding energy peak.
Observation of these neutrinos can tell us a great deal about the mechanism
unless it is distorted en route to the detection.

Attenuation is considered unlikely since neutrinos have a very low rate of
interaction with matter.
\section{Gravitational Effects}
Unlike particle like interactions, which are considered unlikely, the neutrino
interaction with gravitation is small but can accumulate over the flight time
of the burst.  The 1987A neutrino burst observations led to two verifications
of the equivalence principle for neutrinos.  The close arrival time between
neutrinos and light\cite{Longo,Kraus} put bounds on possible violations
of the equivalence principle, including the Shapiro delay\cite{Shapiro} (the accumulated gravitational time delay due to propagation through a gravitational field and the
added distance due to curvature), which
is estimated at 4.8 months for both light and neutrinos.  Angular analysis
of the events suggests the presence of both neutrino and antineutrino events
in the burst which constrains the difference in time delay between matter
and antimatter to essentially the length of the burst, a few seconds.
This neutrino antineutrino comparison is a test of CP violation in general
relativity\cite{LoSecco,Pak}.
\begin{figure}
	\centering
	\includegraphics[width=0.7\linewidth]{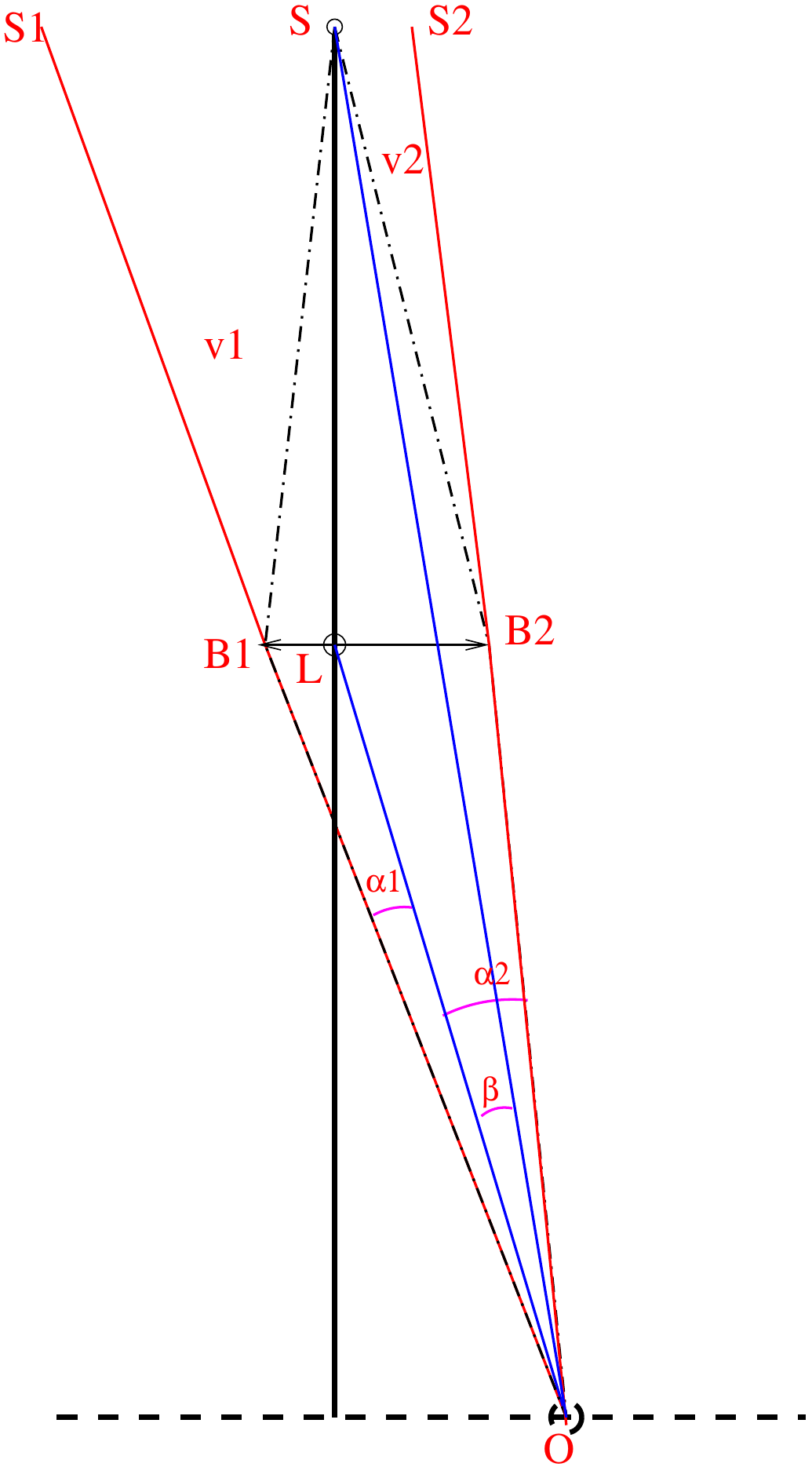}
	\caption{The geometry of a gravitational lens: The source at S when viewed from O appears at the locations S$_{1}$ and S$_{2}$ due to the deflections at B$_{1}$ and B$_{2}$
	caused by the bending of the light by the lens at $L$.
	The angle $SOL$ is $\beta$ and $S_{1}OS_{2}$ is the angle between the two images $\alpha=\alpha_{1}+\alpha_{2}$.}
	\label{resdal}
\end{figure}

\section{Gravitational Lensing and Microlensing}
Gravitational lensing\cite{Klimov,Ref1,Liebes,Ref2} embodies the gravitational
induced curvature in both space
and time.  General relativity predicts that a ray of light
with a distance of closest approach $r$ to star of mass $M$
will be deflected toward the mass $M$ by the angle
$v= \frac{4GM}{c^{2} r}=\frac{K}{r}$.
Multiple images and time delays are possible if multiple light rays from the
same source are directed toward the observer as illustrated in
Figure \ref{resdal}.  An amplification of the
source brightness can occur, as expected from optical lenses.  In the case of
microlensing the images are not spatially resolved and
the evidence may be in the form of variation in the brightness, $L_{1}+L_{2}$,
as the relative positions of the source, the lens and the observer change.
Refsdal\cite{Ref1,Ref2} and Leibes\cite{Liebes} have computed the brightness
amplification and the time delay between the images.

One restriction imposed by Refsdal\cite{Ref1,Ref2} that can
be dropped is that for
neutrinos in most cases the distance of closest approach to
the lens mass, $r$, can be smaller than the radius
of the lens with a suitable redefinition of $M$ since there
will be little attenuation of the neutrino flux.

Figure \ref{resdal} illustrates the lens geometry and defines many of the
variables.
For a pair of images one finds the brightness of the images, $L_{1}$ and $L_{2}$,
\[
L_{1}=\frac{1}{4} (2+\frac{\alpha}{\beta}+\frac{\beta}{\alpha})L_{N}  \]
\[
L_{2}=\frac{1}{4} (-2+\frac{\alpha}{\beta}+\frac{\beta}{\alpha})L_{N}
\]
where $L_{N}$ is the brightness in the absence of a lens.

The time delay is approximately,
\[
\Delta t \approx n D_{L} \alpha \beta c^{-1} (1-\frac{1}{3}\frac{\beta^{2}}{\alpha^{2}})\approx n D_{L} \alpha \beta c^{-1} \approx n D_{L} \alpha_{0} \beta c^{-1}\]
\[\Delta t \approx \frac{16 G}{c^{3}} \frac{\beta}{\alpha_{0}} M =
\frac{8 G}{c^{3}} \frac{\beta}{\theta_{E}} M
\]
where
\[
\alpha=\sqrt{\alpha_{0}^{2}+\beta^{2}} \approx \alpha_{0}(1+\frac{1}{2}\frac{\beta^{2}}{\alpha_{0}^{2}}) \approx \alpha_{0}
\]
\[
\alpha_{0}=\frac{4}{c} \frac{\sqrt{G M}}{\sqrt{n D_{L}}}
\]
and $\beta$ is the angular separation between the lensing object and the direction of the source, $\alpha$ is the angular separation of the two images,
$D_{L}$ is the distance
from the observer to the lens, $n=\frac{D_{S}}{(D_{S}-D_{L})}$, the ratio of the distance from the source to the observer to the distance from the source to the lens.
$n$ always appears in the product
$n D_{L} = \frac{D_{S}D_{L}}{(D_{S}-D_{L})}= \frac{D_{L}}{(D_{S}-D_{L})}D_{S}$
which is the ratio of the observer to lens distance to the lens to source
distance times the distance to the source.
$\alpha_{0}$ is twice the value of the Einstein angle, $\theta_{E}$.

It is clear from the equation for $L_{2}$ that for the second image to be observable one needs $\beta \lesssim \alpha_{0}$.
\[
L_{2}= \frac{1}{4} (-2+\frac{\alpha}{\beta}+\frac{\beta}{\alpha}) L_{N}\rightarrow \lim_{\frac{\beta}{\alpha_{0}} \to \infty}
\frac{1}{4} (-2 + (1+\frac{1}{2}(\frac{\alpha_{0}^{2}}{\beta^{2}})+
1-\frac{1}{2}(\frac{\alpha_{0}^{2}}{\beta^{2}}))L_{N}
\rightarrow 0
\]
and $L_{1}\rightarrow L_{N}$

The time difference between pairs of images, $\Delta t$, is proportional to the ratio $\beta/\alpha_{0}$ and to the lensing mass $M$.

If the two images can not be resolved the observed brightness
is the sum of $L_{1}$ and $L_{2}$.
\[
L_{1}+L_{2}=\frac{1}{2}(\frac{\alpha}{\beta}+\frac{\beta}{\alpha})L_{N}
\]
It is also noteworthy that
\[
L_{1}-L_{2} = L_{N}
\]
and
\[
\frac{L_{1}+L_{2}}{L_{1}-L_{2}}=\frac{1}{2}(\frac{\alpha}{\beta}+\frac{\beta}{\alpha})
\]
,which is a function of only one variable
$\frac{\alpha}{\beta}$ or $\frac{\alpha_{0}}{\beta}$.

From two images one can extract $\frac{\alpha_{0}}{\beta}$ from
the brightness ratio $L_{2}/L_{1}$.  Once $\frac{\alpha_{0}}{\beta}$ is known the lens mass can be extracted from the time delay between the images.
\section{Neutrinos from SN1987A}
The only supernova which has been observed in neutrinos was
SN1987A on February 23, 1987.  Four detectors reported observations of neutrinos on the same day the supernova light
was first seen\cite{KamSN,IMB,Baksan,BaksanE,Baksan1987,MontBlanc}.
These observations are summarized in Table \ref{SNObs}.

The raw information in the Table has not been corrected for the
detection efficiency and energy and trigger bias.  IMB for example could barely detect events below 20 MeV.  
The event
rate in a neutrino detector is proportional to the interacting
mass.  In this case the interacting mass is almost exclusively
the protons (hydrogen).
The Kamiokande\cite{KamSN} and IMB\cite{IMB} detectors are composed of water with 11.2\%
of the mass as hydrogen.  Baksan\cite{Baksan,BaksanE,Baksan1987}
and UNO\cite{MontBlanc} are composed of liquid
scintillator with 14.4\% of the mass as hydrogen.

The number of events recorded in detector $D$ is
\[
N_{D} = N_{A} M_{D} P_{D} \int F(E_{\nu}) \sigma(E_{\nu}) \epsilon_{D}(E_{\nu}) dE_{\nu}
\]
where $N_{A}$ is Avogadro's number, $M_{D}$ is the detector mass in grams, $P_{D}$ is the fraction of the mass in
hydrogen, $\epsilon_{D}(E_{\nu})$ is the energy dependent
efficiency for the detector.  $F(E_{\nu})$ and $\sigma(E_{\nu})$
are the neutrino flux and cross section which are the same for all detectors.

The initial time in the Table was synchronized with a time standard by UNO, IMB and Baksan.  Kamiokande used a computer clock set at computer boot from a watch with a quoted absolute accuracy of $\pm 60$ seconds\cite{KamSN}.

\begin{table}
\scriptsize
\begin{tabular}{|c|c|c|c|c|c|c|}
	\hline
	Name & Mass & Time & Duration & Number of & Mean Energy & Energy Range\\
	& Metric tons & UTC & seconds & Events & MeV & MeV \\
	\hline
	UNO & 90 & 02:52:36.79 & 7.01 & 5 & 8.4 & 7-11 \\
	\hline
	Kamiokande & 2140 & 07:35:35 & 12.439 & 9+3=12 &  14.7 & 6.3-35.4\\
	\hline
	IMB & 5000 & 07:35:41.37 & 5.59 & 6+2=8 &  32.5 & 20-40 \\
	\hline
	Baksan & 200 & 07:36:11.818 & 9.099 & 3+2=5 & 18.1 & 12-23.3\\
	\hline
\end{tabular}
\normalsize
\caption{\label{SNObs}Summary of the reported neutrino observations of SN1987A.  The mean energy is a simple average of the energy of the reported events.  It is not corrected for
efficiency.  The mean energy gives a rough idea of the
portion of the energy spectrum sampled by that detector.
The detectors have different capabilities and can only be compared after corrections for inefficiencies and other biases
are accounted for.}
\end{table}
\section{Observational Evidence}
Observational evidence for gravitational lensing of neutrino bursts is limited,
since as of this writing only one burst has been observed.
The 4 reported observations in Table \ref{SNObs} have many incompatibilities.
It is unlikely that gravitational lensing would be able to explain them.

All of the observations were made from Earth.  The neutrino
source was in the Large Magellanic Cloud about 50 kpc away.
The maximum angular separation between the observations as
viewed from the source is $8 \times 10^{-15}$ radians so
all observations of the supernova neutrinos saw the same thing.
Differences between the observations can occur because of
differences in the sensitivity of each detector, instrumental
effects and problems, or background events unrelated to the
supernova.  For example the IMB detector had a 14\%
efficiency for triggering the recording of an event at 20 MeV
and lower efficiency at lower energies.  It is unlikely IMB could have seen the
events reported by UNO, all below 11 MeV, even though IMB had a proton mass
43 times larger.  Kamiokande spans the broadest energy range with a good
sensitivity overlap with all detectors. 

One should also be careful in over estimating the precision of the measurements.
Neutrino observations are made quanta by quanta.  One wants to understand
the underlying distributions based on a limited number of samples.  For
example, the accuracy of time differences for events from the same
detector are about 1 millisecond.  But the rate of sampling, in places, is
well below this, about one event per second.  So time differences between images
only have a physical significance to a second or more.

Lensing manifests itself in three ways.  There can be multiple images.
The images can be brighter than one which has not
been lensed.  The different images can have time
delays between them due to the different paths taken to the
observer.  In the case of simply two images there are four
observables, the angle $\alpha$ between the images, the brightness $L_{1}$ and
$L_{2}$ of each image and the time delay
between the images.

\begin{figure}
	\centering
	\includegraphics[width=0.7\linewidth]{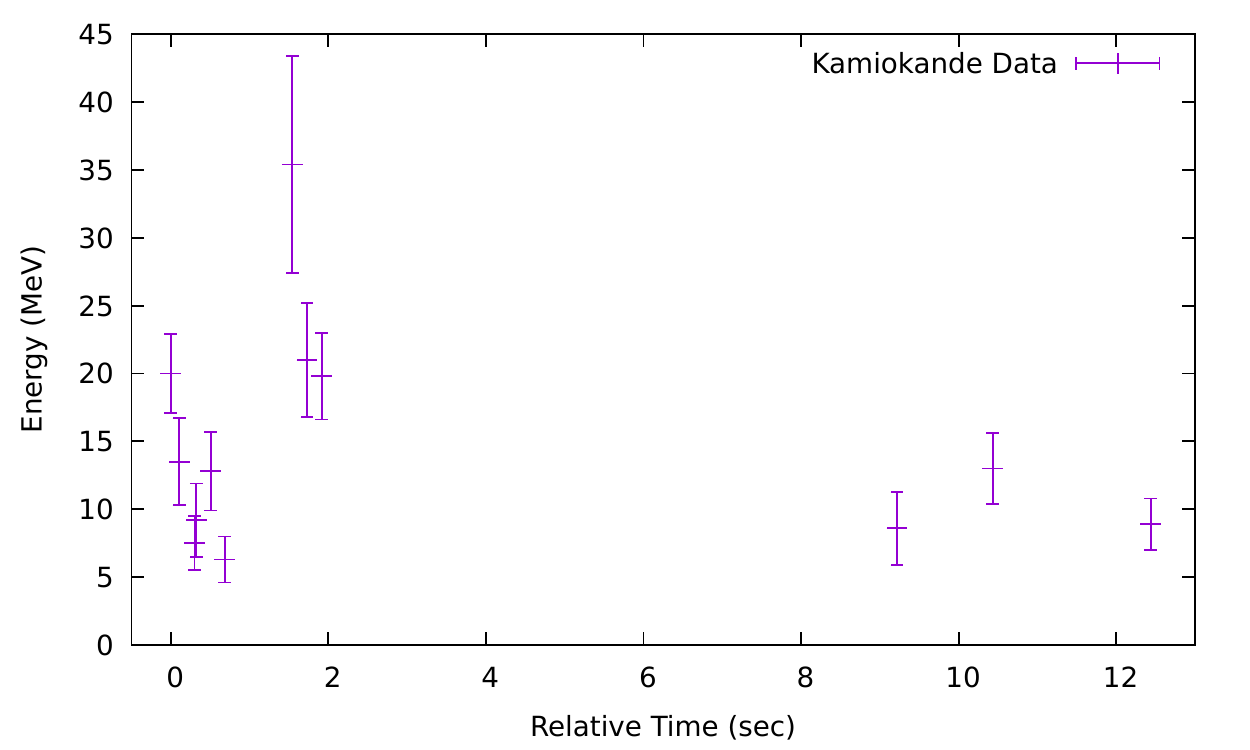}\\
	\includegraphics[width=0.7\linewidth]{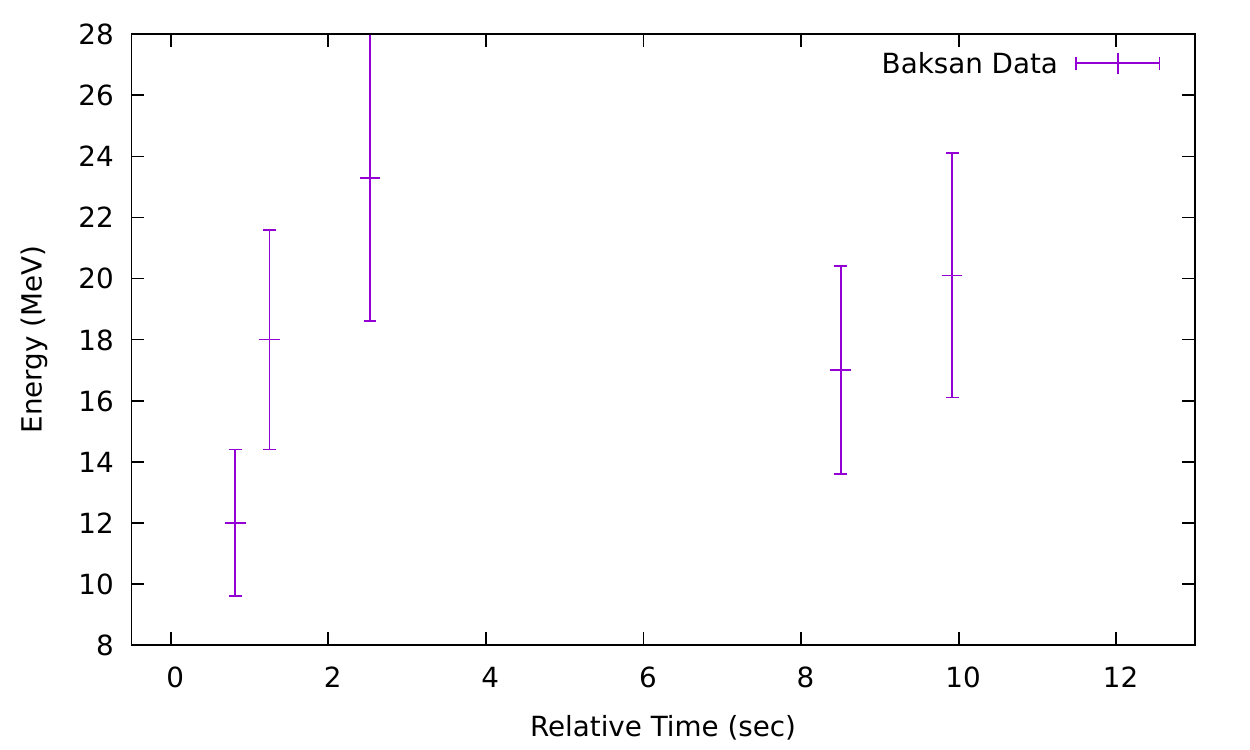}
\caption{The Kamiokande neutrino events (top) and Baksan (bottom) from SN1987A 
  plotted as event energy as a function of relative time. There is a notable gap
  in both data samples.  The relative time between the plots
is arbitrary.  It is not the nominal time listed in Table \ref{SNObs}}
	\label{kamsn1987a}
\end{figure}

The detection reaction for most of these events
was electron antineutrino charged current with a target proton,
producing a positron and a neutron,
$\bar{\nu}_{e} + P \rightarrow e^{+} + N$.
This reaction can provide good energy resolution, even in the absence of
observing the neutron, but it does not provide
any appreciable directionality.  (Some experiments have tried to determine the
direction of the neutrino by measuring the distance between the source of the
positron and the neutron capture.  Angular sensitivities of about 8$^{\circ}$
are possible with very large samples.\cite{Caden}  None of the experiments
in Table \ref{SNObs} used this method.)  The lack of directional sensitivity
makes the observation of the angle $\alpha$ between any two images unlikely.

The lens is geometrical so it is expected
to be achromatic.  It will not distort the neutrino spectrum
by bending different neutrino energies to different angles.
In the case of light, the images may have different spectra,
since the paths through the interstellar medium will differ
and the two paths may encounter different materials.

One can use the time distribution, such as time clustering,
to identify gravitational lensing of the neutrino burst.
Brightness differences between these clusters can also help
identify a lensing event.  If the time clusters have distinct
spectra, except for brightness differences it is less likely
they manifest multiple observations of the same event.  A timing difference can
only be observed if both lensed images are bright enough to be observed.

The recorded times of the events, in particular the time of events from one
detector relative to other events from the same detector, are well measured.
In principle the relative
time of events from standard time synchronized experiments
should also have good relative time measurements.  Two timing
issues are apparent from Table \ref{SNObs}.  UNO reported a signal
4 hours and 43 minutes before IMB and Baksan.  IMB reported
a signal about 30.4 seconds before Baksan.  The difference in
nominal time of 6 seconds between Kamiokande and IMB has no significance given
Kamiokande's $\pm 1$ minute error on the
absolute time.  As mentioned above, except for instrumental differences,
all detectors should see the same thing.  Lensing can not explain the
differences in start times.

The Kamiokande data has a 7.3 second gap starting at
1.915 seconds, see Figure \ref{kamsn1987a}.  The full burst reported by Kamiokande has a duration of 12.439 seconds.  The data could be understood
as two bursts.  The first lasting 1.915 seconds contained 9 events and the second
lasting 3.22 seconds had 3 events.  The time difference between these
two pulses is 9.219 seconds.  Using the event counts as an estimate of the
brightness suggests $L2/L1 = 3/9 =\frac{1}{3}$.  Can gravitational lensing
explain the time structure and brightness ratio?

\begin{figure}
	\centering
	\includegraphics[width=0.7\linewidth]{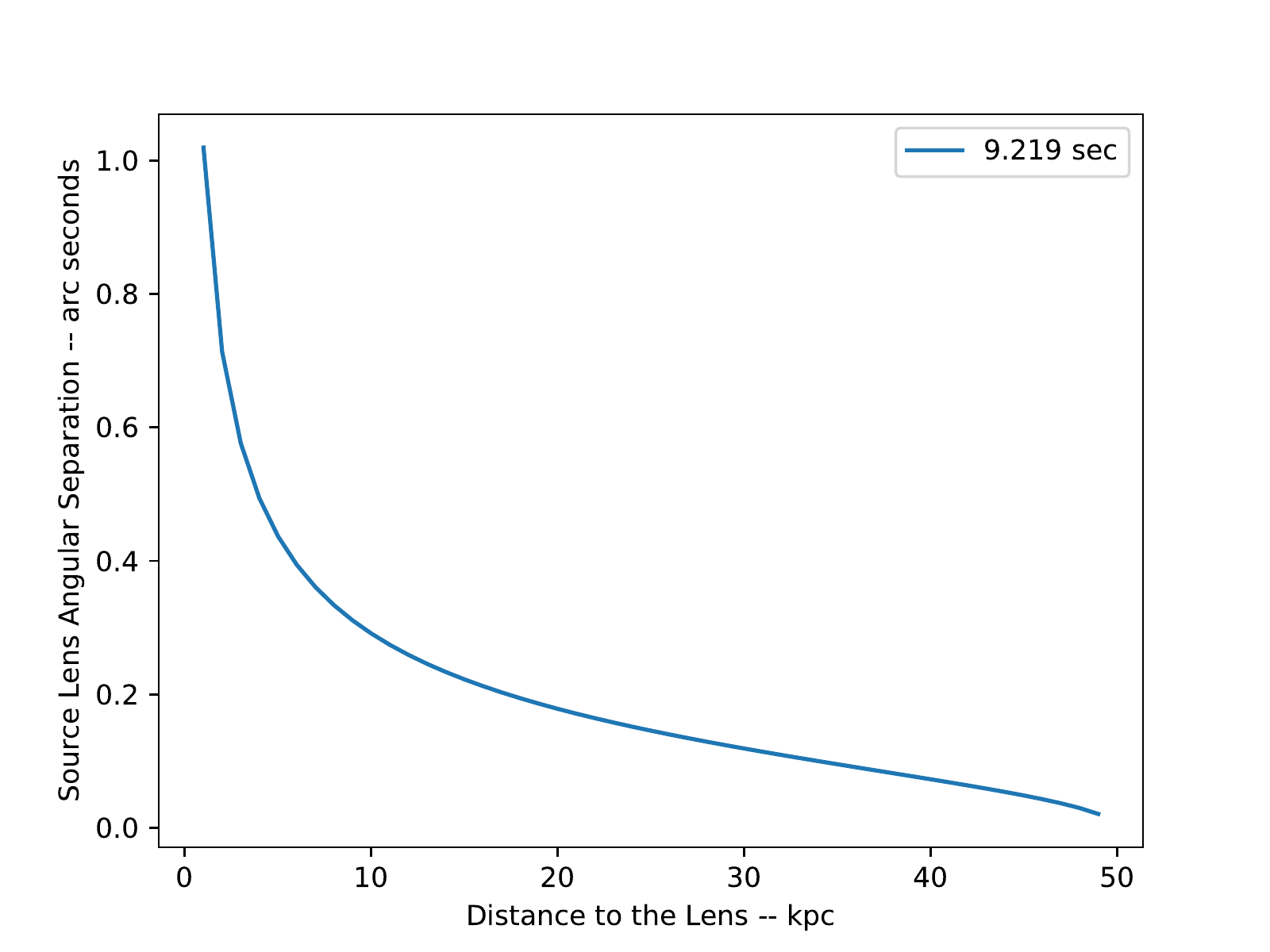}
	\caption{The angle between the lens and the source, $\beta$ in seconds
          of arc versus the distance to the lens for a solution for $M$
          and $\alpha_{0}/\beta$ that
          fits the Kamiokande time and brightness structure.}
	\label{BetavsaB}
\end{figure}

From the Kamiokande pulse brightness ratio $L_{2}/L_{1} = \frac{1}{3}$
one computes  $\alpha_{0} = 3.59 \beta$.  From the value of the time
delay of $\Delta t = 9.219$ and $\frac{\alpha_{0}}{\beta}$ one calculates the
lens mass at about 414,000 solar masses.  A scan of the lens distance from
1 to 49 kpc gives an angular separation between the two images, $\alpha$ from
3.8 arc seconds at the shortest distance to 0.08 at 49 kpc.  The lens source
angular separation ($\beta$) as a function of the distance to the lens is
shown in Figure \ref{BetavsaB}.  It drops from 1.02 arc seconds near the observer to 0.03 to 0.02 near the source.

Comparison with other experiments is difficult since Kamiokande did not have
a synchronized clock.  Baksan had a synchronized signal but it was about 30.448
after the synchronized IMB signal.  This Baksan time shift
will be revisited in the next section.  One can analyze
the Baksan signal itself for compatibility with the lensing
hypothesis.

The Baksan signal of 5 events has a structure of 3 early
events over 1.71 seconds followed by 2 events in 1.412 seconds,
Figure \ref{kamsn1987a}.  The gap between
these bunches is 7.687 seconds.  The ratio of the energy in each pulse is
crudely the ratio of the number of events which is 0.667$\pm$0.609 which
is not in conflict with the Kamiokande observations.
So Baksan may have seen the first pulse at 1.3 to 1.5 times the nominal
brightness and a second pulse suppressed to 0.3 to 0.5 the nominal brightness.
While not a strong confirmation since the brightness ratio is very weak,
the Baksan results support the two pulse gravitational lens interpretation.

IMB also had a time structure in its event sample.
Of the 8 events observed overall 5 occurred in the first 1.57
seconds of the burst, compatible with the time scale of the
first 9 Kamiokande events.  The last 2 IMB events occurred
in 0.58 seconds 5.01 seconds after the initial event.
The 6'th IMB event occurred at 2.69 seconds which is outside the 2 second
duration of the ``first'' pulse observed
at Kamiokande and Baksan.  The ratio of late to early events,
2/6 is compatible with the other evidence.  The second pulse
seems to be too early to fit the lensing picture supported by
the other observations.  The shorter time gap
suggests a smaller lensing mass but also casts doubt
on the lens hypothesis.

There is no known star cluster in the direction of the LMC to provide the mass.
The source lens alignment, Figure \ref{BetavsaB} seems rather unlikely.
The standard estimate\cite{Griest} of the galactic ``optical depth'' for
lensing a source in the LMC is about $10^{-6}$ independent of the lensing mass.

In terms of the physics of supernovae the lensing amplified
the energy recorded by Kamiokande by a factor of
$\frac{L_{1}+L_{2}}{L_{1}-L_{2}}\approx$2.0 since the
first pulse is brightened by a factor of 1.5 and the second pulse adds in
another 0.5.  So the reported energy in electron antineutrinos of
$8 \times 10^{52}$ ergs\cite{KamSN} is an over estimate.

\section{Alternate Explanations}
A number of ideas have been put forward to explain
the UNO event time.  While UNO did not report a signal at the
7:35:41 time reported for other neutrino observations, they
have reported coincidences with a gravitational wave bar detector\cite{Amaldi}.  The absence
of a neutrino signal in UNO at
7:35:41 is not unexpected since its mass is small, 2.2 times smaller than
Baksan, so it would be expected to have only 2.25 events at that time.

It has been suggested that the 30 second delay of Baksan
with respect to other observations is due to a systematic 30 second advance added to the time standard in the USSR\cite{Learned}.

Oyama\cite{Yuichi} has suggested that the 7.3 second gap
in the Kamiokande data is due to an instrumentation inefficiency that made
data collection impossible in that interval.

The most popular opinion is that UNO is unexplained but not
identified with SN1987A and that the time structures and energy differences of
the other observations are due to instrumental differences and
statistical fluctuations in small data samples.
The late time distribution of neutrino events is considered
indicative of neutrino cooling of the source.
\section{Gravitational Lensing of Optical Supernovae}
Gravitational lensings of the optical components of supernovae
have been seen\cite{Refsdal}.  Four images of a supernova at
z=1.49 were created by a lensing galaxy, J1149.5+2223, at z=0.54.  The lensing
manifested itself as four images observed
in April 2015.  A fifth image appeared in December 2015 at
a predicted location.

Goobar {\em et al.}\cite{Goobar2017}
has reported multiple images from a Type 1a supernova at
z=0.409 at an amplification of more than 50, lensed by a galaxy at z=0.216.
Four images were resolved.

Several additional supernovae have been lensed\cite{Goobar,Patel,Rodney,Chornock,Quimby},
manifesting higher brightness than expected.
\section{Conclusion}
Gravitational lensing can distort the observation of any distant object
including a supernova neutrino burst.
It is unlikely but should not be overlooked if the data shows noteworthy
departures from expectations such as multiple pulses or unexpectedly large
signals.  Redundancy of observations is helpful in avoiding
misleading results from statistical fluctuations.

The distribution of neutrinos from SN1987A have a time and
brightness structure expected from a lensing event.  The
mass and angular alignment make this unlikely but perhaps
optical confirmation of such a lens is still possible.
A lens will influence starlight coming from behind it if the alignment
is appropriate.  Observing a time delay as suggested by the neutrino burst
requires some time structure in the source of tens of seconds or less.
It also requires detecting equipment sensitive to short time scale variation
in the brightness.  On the other hand an optical detection may be able
to measure the angular separation of the images and the ratio of their
brightness.  It is likely that alignments have changed over the decades
and one must search back through the historical record.  Multiple images of
a single star in the LMC are hard to distinguish from multiple stars
in the LMC especially since the images will not have the same apparent
luminosity.
Lensing of starlight is difficult since, unless the alignment with the lens
changes, there is no way to tell that the image has been displaced from its flat
space location.  Image distortion or brightness variation
caused by microlensing would help to identify the presence of
a lens.

The lensing hypothesis for the observed SN1987A neutrino signal would reduce
the total energy
estimate derived from the observation since the brightness would have
been enhanced.  On the other hand the average neutrino power in the pulse
would increase since the duration of the pulse
would be shortened to 2-3 seconds.  Not all observations support this
interpretation.
\section{Acknowledgments}
I am grateful to Kate Scholberg who convinced me that one need not consider
SN1987A as typical.
John Learned, Bob Svoboda and Ralph Becker-Szendy have contributed important
ideas to a discussion initiated by reference \cite{Yuichi}.
Erica Caden filled me in on the status of direction finding with
the $\bar{\nu}_{e} P$ reaction, reference \cite{Caden}.
Jim Rich had many suggestions after reviewing an early version of the
manuscript, including updating to a more modern notation.

Note added in proof:
Shantanu Desai has pointed out that the lensing hypothesis for the
SN1987A neutrino signal had been proposed in 1987, Barrow and
Subramanian \cite{Nature1987}.  The current work extends the analysis by
including fits to three of the observations and showing the mass estimate is
independent of an assumed distance to the lens.  I thank him for bringing
reference \cite{Nature1987} to my attention.

\end{document}